\pdfoutput=1
\documentclass[twocolumn,showpacs,preprintnumbers,amsmath,amssymb,pre]{revtex4-1}
\usepackage[utf8]{inputenc}
\usepackage{amsopn}
\usepackage{bm,bbm}
\usepackage{subfig}
\usepackage{amsmath}
\usepackage{amssymb}
\usepackage{multirow}
\usepackage{hyperref}
\usepackage{graphicx}
\newcommand{\ket}[1]{\ensuremath{|#1\rangle}}

\newcommand{\braket}[2]{\ensuremath{\langle{#1}|{#2}\rangle}}

\newcommand{\1}{{\rm 1\hspace{-0.9mm}l}}
\newcommand{\Id}{\1}
\begin{document}

\title{Quantum Prisoner's Dilemma game on hypergraph networks}

\author{{\L}ukasz Pawela}
\email{lukasz.pawela@gmail.com}
\affiliation{Institute of Theoretical and Applied Informatics, Polish Academy
of Sciences, Ba{\l}tycka 5, 44-100 Gliwice, Poland}

\author{Jan S{\l}adkowski}
\email{jan.sladkowski@us.edu.pl}
\affiliation{Institute of Physics, University of Silesia,
Uniwersytecka 4, 40-007 Katowice, Poland}

\begin{abstract}
We study the possible advantages of adopting of quantum strategies in multi-player evolutionary games.
We base our study on the three-player Prisoner's Dilemma (PD) game. In order to model the simultaneous
interaction between three agents we use hypergraphs and hypergraph networks. In particular, we study
two types of networks: a random network and a SF-like network. The obtained results show that
in the case of a three player game on a hypergraph network, quantum strategies 
are not necessarily stochastically stable strategies. In some cases, the 
defection strategy can be as good
as a quantum one.
\end{abstract}

\date{27 February 2012}
\pacs{03.67.-a, 02.50.Le}

\maketitle

\section{Introduction}
Game theory is a branch of mathematics broadly applied in a great number of fields, from biology to social
sciences and economics.  A great deal of effort has gone into the study of evolutionary
games on graphs, which was initiated by the work  of Nowak and May \cite{nowak1992evolutionary}. Since their work was published,
a lot of effort was put into studying the problem \cite{szabo2007evolutionary}.

Quantum game theory \cite{piotrowski_invitation_2002} allows the agents to use quantum strategies. The set of quantum strategies is much
larger than a classical one; hence it offers possibility for much more diverse behavior of agents in the network.
It has been shown that if only one player is aware of the quantum nature of the system, he/she will never lose
in some types of games \cite{eisert1999quantum}. Recently, it has been demonstrated that a player can cheat by appending
additional qubits to the quantum system \cite{miszczak2011qubit}.

Combining evolutionary games and quantum game theory, has resulted in absorbing results \cite{abbot}. In some
cases the quantum strategies can dominate the entire network, infecting it effectively. In our work we
like to focus on introducing additional strategies which the agents can use, since in the multi-player
case there exists a Pareto Optimal Nash Equilibrium for the Prisoner's Dilemma game \cite{du2002entanglement}.
Moreover, the PD game is interesting to study, because it was realized experimentally \cite{du2002experimental}.

A hypergraph is a concept that generalizes the concept of a graph by allowing edges to connect more than two nodes at once. This concept can be applied for systems with evolution described by the extended spin-$1/2$ chain~\cite{lou2004thermal}. Based on this, the notion of hypergraphs was proven to be a useful tool for assuring controllability of multipartite quantum systems~\cite{puchala2012local}.

This paper is organized as follows: In Section \ref{sec:qgry} a short review of quantum game theory is given and in Section \ref{sec:net}  the types of 3-hypergraph networks
used in simulations are described. Section \ref{sec:game} introduces the three-player
Prisoner's Dilemma game. In Section \ref{sec:setup} the simulation setup is described.
Section \ref{sec:res} contains results obtained from computer simulations and their
discussion. Finally, in Section \ref{sec:conc} the final conclusions are drawn.

\section{Quantum game theory} \label{sec:qgry} Games that admit the player to use the peculiarities  of quantum phenomena are referred to as quantum games \cite{piotrowski_invitation_2002, Guo, eisert1999quantum}. Of course, they are games  in the "classical" sense.  Actually, any quantum system which can be manipulated  by at least one party and where
the utility of the moves can be reasonably quantified, may be
conceived as a quantum game. To be more specific, a {\it two-player quantum game}\/
$\Gamma=({\cal H},\rho,S_A,S_B,P_A,P_B)$ is completely specified
by the underlying Hilbert space ${\cal H}$ of the physical system,
the initial state $\rho\in {\cal S}({\cal H})$, where ${\cal
S}({\cal H})$ is the associated state space, the sets $S_A$ and
$S_B$ of permissible quantum operations of the two players, and
the { pay-off (utility) functions}\/ $P_A$ and $P_B$, which
specify the pay-off for each player. A {\it  quantum strategy}\/ $s_A\in S_A$, $s_B\in S_B$ is a
quantum operation, that is, a completely positive trace-preserving
map mapping the state space on itself. The quantum game's
definition may also include certain additional rules, such as the
order of the implementation of the respective quantum strategies. The generalization to the multi-player case is straightforward.  Schematically we have:
$$
\rho \mapsto (s_{A},s_{B}) \mapsto  \sigma \Rightarrow (P_{A},
P_{B}).$$
The following concepts  are commonly used in the context of quantum game theory. These definitions are fully analogous to the
corresponding definitions in "classical" game theory. A quantum strategy  $s_A$ is called {\it dominant
strategy}\/ of Alice if
\begin{eqnarray}
        P_A(s_A,s_B')
        &\geq&
        P_A(s_A',s_B')
\end{eqnarray}
for all $s_A'\in S_A$, $s_B'\in S_B$. Analogously we can define a
dominant strategy for Bob. A pair $(s_A,s_B)$ is said to be an
{\it equilibrium in dominant strategies}\/ if $s_A$ and $s_B$ are
the players' respective dominant strategies. A combination of
strategies $(s_A,s_B)$ is called a {\it Nash equilibrium}\/ if
\begin{eqnarray}
        P_A(s_A,s_B)&\geq& P_A(s_A',s_B),\\
        P_B(s_A,s_B)&\geq& P_B(s_A,s_B') .
\end{eqnarray}
A pair of strategies $(s_A, s_B)$ is called {\it Pareto
optimal}\/, if it is not possible to increase one player's pay-off
without lessening the pay-off of the other player. A solution in
dominant strategies is the strongest solution concept for a
non-zero sum game. In the Prisoner's Dilemma \cite{eisert1999quantum,piotrowski_invitation_2002}
:

$$ \begin{array}{c|cc}
     & \mbox{Bob}: C & \mbox{Bob}: D \\
    \hline
    \mbox{Alice}: C & (3,3) & (0,5) \\
    \mbox{Alice}: D & (5,0) & (1,1)
  \end{array}
$$
(the numbers in parentheses represent the row (Alice) and column
(Bob) player's payoffs, respectively). Defection is the dominant
strategy, as it is favorable regardless what strategy the other
party chooses. In a Nash equilibrium  neither player
has a motivation to unilaterally alter his/her strategy, as this action will not imcrease his/her pay-off.
Given that the other player will stick to the strategy
corresponding to the equilibrium, the best result is achieved by
also playing the equilibrium solution. The concept of Nash
equilibrium is therefore of paramount importance. However, it is only an acceptable solution
concept if the Nash equilibrium is not unique. For games with
multiple equilibria we have to find a way to eliminate all but one
of the Nash equilibria. A Nash equilibrium is not necessarily
efficient. We say that an equilibrium is  Pareto optimal if there
is no other outcome which would make all players better off. Up
to now a lot of papers on quantum games have been
published and some application outside the field of physics have also been discussed \cite{piotrowski_newcomb_2003,arfi,sladkowski_giffen_2003,PSS_2003,Riccardo_2009,Busemeyer_2006,PS_2005}.

\section{Hypergraphs and hypergraph networks} \label{sec:net}
We assume a hypergraph \cite{hypergraph} network $H(X,E)$ where $X$ is a set of nodes
and $E$ is a set of non-empty subsets of $X$, $E\subseteq 2^X$. Elements of $E$
are the hyperedges of $H$. We keep within the boundaries of the case when every subset
 of $X$, $A\in E$ satisfies $|A|=3$, i.e. every edge of the hypergraph
connects three nodes exactly. Hereafter we will refer to this structure as a 3-hypergraph.
We set $N=|X|$ -- the total number of agents.

We construct two types of networks: a random network, in which all hyperedges connect
random nodes and a SF-like \cite{barabasi1999emergence} network. We set the number of hyperedges in the random
case to $|E| = 10000$. The SF-like network is constructed in the following way:
First, a network of $m_0 \ll N$ all connected nodes is created. Then a new node with $m < m_0$ links
is added to the network. For each of the $m$ links, a pair of unique nodes is chosen from the existing
network and a new hyperedge is added. The probability of a node $i$ being chosen is given by:
	\begin{equation}
		p_{sf}(i) = \frac{k_i}{\sum_{j \in X} k_j},
	\end{equation}
where $k$ is the degree of a node. This procedure is repeated until the number of nodes of the network
reaches N.

\section{Three-player PD game} \label{sec:game}

The classical Prisoner's Dilemma game is as follows: two players can either cooperate ($C$) or defect ($D$).
When they both cooperate, each receives a payoff of $3$. On the other hand, when they both defect, each receives
a payoff of $1$. When one defects, he/she receives a payoff of $5$, while the other gets $0$.

This approach can be extended to a greater number of players. In the three-player case,
the payoff matrix is shown in Table \ref{tab:payoff}.
	\begin{table}[!htbp]
		\centering\begin{tabular}{cc}
			\underline{Charlie $C$} & \underline{Charlie $D$} \\
			\begin{tabular}{cccc}
					&  & \multicolumn{2}{c}{Bob} \\
					& \multicolumn{1}{c|}{} & $C$ & $D$ \\ \cline{2-4}
					\multirow{2}{*}{Alice} & \multicolumn{1}{c|}{$C$} & $(6,6,6)$ & $(3,9,3)$ \\
					& \multicolumn{1}{c|}{$D$} & $(9,3,3)$ & $(5,5,0)$ \\
			\end{tabular}
			&
			\begin{tabular}{cccc}
					&  & \multicolumn{2}{c}{Bob} \\
					& \multicolumn{1}{c|}{} & $C$ & $D$ \\ \cline{2-4}
					\multirow{2}{*}{Alice} & \multicolumn{1}{c|}{$C$} & $(3,3,9)$ & $(0,5,5)$ \\
					& \multicolumn{1}{c|}{$D$} & $(5,0,5)$ & $(1,1,1)$ \\
			\end{tabular}
		\end{tabular}	
		\caption{The payoff matrix of the three-player PD game (after \cite{chappell2011analyzing}). The first entry is the payoff of Alice, the
		second denotes the payoff of Bob and the third represents the payoff of Charlie.}\label{tab:payoff}
	\end{table}
We can see that every player is better off defecting than cooperating no matter what the other players do.
In terms of game theory, $(D,D,D)$ is the unique Nash equilibrium of the game. If any one player deviates
from this strategy, he will receive a lower payoff. On the other, we can see that the strategy profile$(C,C,C)$
can yield a higher payoff than $(D,D,D)$. In terms of game theory this profile is Pareto Optimal. In our
case the players are rational and the game will end in $(D,D,D)$, not $(C,C,C)$; hence the dilemma.

\begin{figure}[!htbp]
		\centering\includegraphics{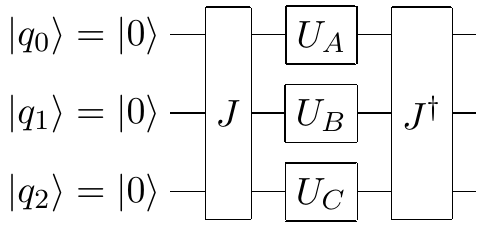}
		\caption{Quantum circuit for the three-player PD game. $U_A$, $U_B$, $U_C$ are the strategies of Alice, Bob
				and Charlie respectively.}\label{fig:circuit}
	\end{figure}

In the quantum case the setup is as follows. Each player is sent a qubit and can locally operate
on it, using any unitary operator $U \in SU(2)$. The initial state of the system is entangled:
	\begin{equation}
		\ket{\psi} = J\ket{000},
	\end{equation}
where $J$ is the entangling operator \cite{multiqubit_entangling}:
	\begin{equation}
		J = \frac{1}{\sqrt{2}} \left(\Id^{\otimes N} + i\sigma_x^{\otimes N}\right).
	\end{equation}
The quantum circuit for the game is shown in Figure \ref{fig:circuit}.
After the players have applied their respective strategies, the untangling gate, $J^\dagger$, is applied to the system,
hence the final state of the game is
	\begin{equation}
		\ket{\psi_f} = J^\dagger(U_A \otimes U_B \otimes U_C)J\ket{000},
	\end{equation}
where $U_A,U_B,U_C$ are the players strategies.
The payoff of the first player (Alice) amounts to:
	\begin{equation}
		\$_A = \sum_{i,j,k \in \{0,1\}} p_{ijk} |\braket{\psi_f}{ijk}|^2,
	\end{equation}
where $p_{ijk}$ are numbers corresponding to the possible classical payoffs of Alice, defined in Table \ref{tab:payoff}.

\section{Simulations} \label{sec:setup}
We assume an initial population of 2500 agents, located at the nodes of a hypergraph. The SF-like network is
constructed with initial size $m_0=3$, and the number of links of each new node is $m=2$. Throughout all simulations the network topology remains static. The set
of allowed strategies is as follows \cite{abbot}:
	\begin{equation}
		S = \{C, D, H, Q\},
	\end{equation}
where the unitary operators corresponding to each of the strategies take the form of::
	\begin{equation}
	\begin{split}
		C &= \left( \begin{array}{cc}
			1&0\\
			0&1
		\end{array} \right),\;
		\phantom{-\frac{1}{\sqrt{2}}} D = \left( \begin{array}{cc}
				0&1\\
				1&0
			\end{array} \right),\\
		H &= \frac{1}{\sqrt{2}}\left( \begin{array}{cc}
				1&1\\
				1&-1
			\end{array} \right),\;
		Q = \left( \begin{array}{cc}
				i&0\\
				0&-i
			\end{array} \right).
	\end{split}
	\end{equation}
In the two-player case, the strategy profile ($Q$,$Q$) is a Nash Equilibrium of the game and the strategy $H$ introduces a \emph{miracle move} when an agent uses it against other's classical strategy \cite{eisert1999quantum}.
These strategies are randomly assigned to agents in the network in such a way that the initial
fractions of strategies $C$, $D$, $H$, $Q$ are 49\%, 49\%, 1\%, 1\% respectively.

Next, we introduce an additional strategy $\Sigma$, defined as:
	\begin{equation}
		\Sigma = i\sigma_y=\left( \begin{array}{cc}
					0&1\\
					-1&0
				\end{array} \right).
	\end{equation}
The strategy profile ($\Sigma$, $\Sigma$, $\Sigma$) is Pareto Optimal and a Nash Equilibrium \cite{du2002entanglement}.
We assign the strategies $C$, $D$, $H$, $Q$, $\Sigma$ with frequencies 48\%, 48\%, 2\%, 1\%, 1\%.

Finally, we do not assign strategies randomly, but choose to allocate the strategy $Q$ in the first
case and $\Sigma$ in the second one to nodes with the highest degree.

The evolution of strategies of agents is as follows. Initially, each agent is assigned a strategy based on the rules described in the former two paragraphs. Next, agent $i$ plays an entangled quantum game with all other agents forming a hyperedge with him/her. This is repeated for all hyperedges $e \in E$ such that $i\in e$. The total payoff of agent $i$, $F_i$ is obtained by accumulating all the payoffs from these games. After that, the agent $i$ chooses one of these agents randomly, denoted $j$, and imitate its strategy with probability $p_i$~\cite{PhysRevE.72.056128},
\begin{equation}
	p_i=\left\{
	\begin{array}{cc}
	\frac{F_j - F_i}{\alpha \mathrm{max}(k_i,k_j)}, & F_j > F_i \\
	0, & \mathrm{otherwise}
	\end{array}
	\right.,
\end{equation}
where $\alpha = \max ( \{p_{ijk} \} )$.

The PD game is played by all agents on both networks. We study the impact of the value of the parameter
$T$ (moral hazard) on the final state of the population. This parameter is defined as the first players
payoff when other players use the $C$ strategy. Its interpretation is as follows.
Suppose the prisoners had a chance to discuss a strategy. It is evident that they should decide for a Pareto Optimal
profile $(C,C,C)$. However, if Alice decides to defect, she receives a higher payoff. Thus this parameter measures, how much Alice is tempted to betray the other prisoners.

The game is played for 10000 generations
and the last 1000 results are stored. Average frequencies of strategies are used as the final results.
If a population does not change for 500 generations, the state is considered to be an equilibrium state of the system.

\section{Results and discussion} \label{sec:res}
In all cases, we show results for T starting from 5, despite the fact that 
the game becomes a Prisoner's Dilemma when $T >6$. We have done this to show 
the behavior of the fractions of strategies near the transition to the PD game.

\begin{figure}[!htbp]
		\centering
		\subfloat[Random network]{\label{fig:rn4s}\includegraphics{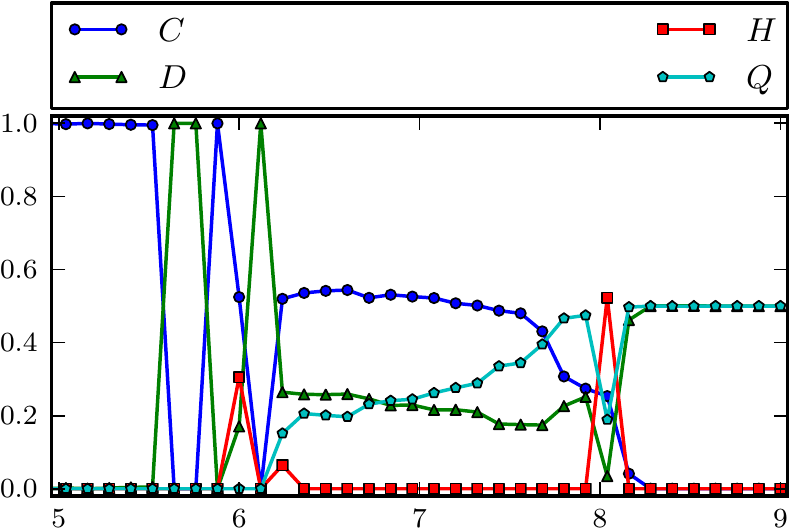}}\\
		\subfloat[SF-like network]{\label{fig:sf4s}\includegraphics{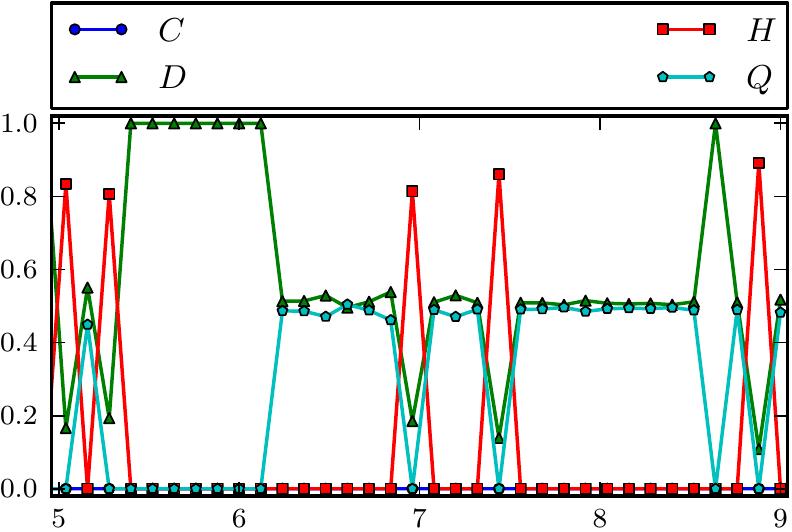}}
		\caption{Results for PD on hypergraph networks, 4 strategies, strategies assigned at random, according to weights.}\label{fig:4s}
	\end{figure}

In the case with four possible strategies, the results of computer simulations are depicted in Figure \ref{fig:4s}.
Figure \ref{fig:rn4s} shows the results for a random network, whereas the results for the SF-like network are shown
in Figure \ref{fig:sf4s}. In the case of a random network, we see that strategy $C$ is the dominant one, until $T=5.64$, when
the network starts shifting between strategies $C$ and $D$. It settles down at $T=6$, where about half the agents
use strategy $C$. As $T$ increases, strategies $C$ and $D$ slowly lose their significance in favour of strategy $Q$.
For $T > 8$ the system reaches another equlibrium state, where strategies $D$ and $Q$ have the same frequency.

In the case of a SF-like network, the agents prefer the $D$ strategy, which almost never reaches zero frequency.
Again, for $T>6$ we have an increase of significance of the quantum strategy $Q$. Although there are some oscillations
of the fraction of strategies as $T$ increases, again strategies $D$ and $Q$ have been adopted by approximately the same
fraction of agents. On the basis of the presented figures as well as above discussion it may be inferred that the change
of type of the network significantly decreases the importance of strategy $C$, but does not have a great impact on strategies $D$ and $Q$.

\begin{figure}[!htbp]
		\centering
		\subfloat[Random network]{\label{fig:rn5s}\includegraphics{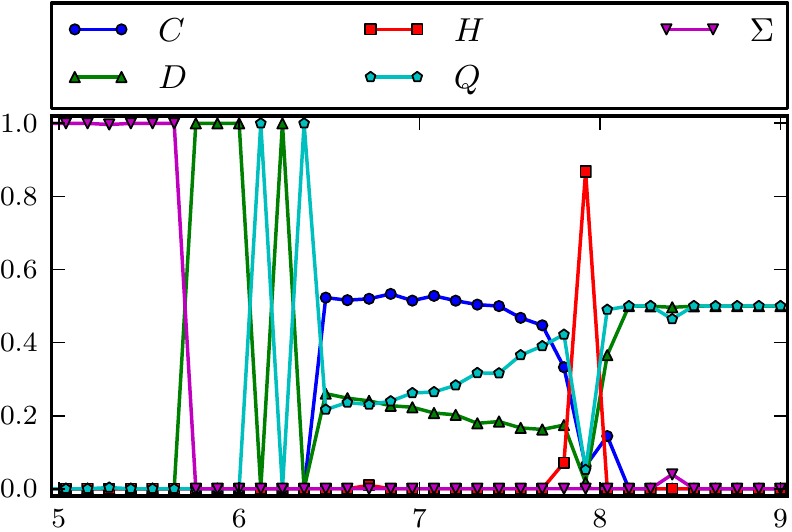}}\\
		\subfloat[SF-like network]{\label{fig:sf5s}\includegraphics{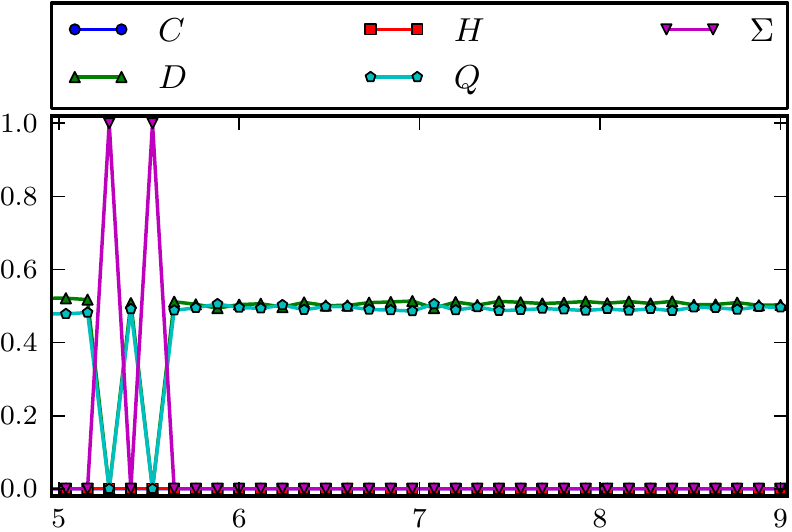}}
		\caption{Results for PD on hypergraph networks, 5 strategies, strategies assigned at random, according to weights.}\label{fig:5s}
	\end{figure}

Figure \ref{fig:5s} illustrates the results obtained for five possible strategies. Figure \ref{fig:rn5s} illustrates the
results for a random network, and Figure \ref{fig:sf5s} shows the results for the SF-like network.
The examination of Figure \ref{fig:rn5s} reveals that it has the same character as the Figure \ref{fig:rn4s}, except
that for $T<6$ the dominant strategy is $\Sigma$ not $C$. At around $T=6$, the network shifts from $\Sigma$
dominated to a network with three possible strategies: $C$, $D$, $Q$. As $T$ increases, strategies $C$ and
$D$ lose their significance in favor of $Q$. At around $T=8$, there is another shift in strategies, and the
fraction of strategy $C$ decreases to zero, and strategies $D$ and $Q$ are used by equal fraction of agents.
	
On examining the SF-like network, we again perceive behaviour similar to the four strategy case, but with much less oscillations. Again for $T>6$
it is observed that strategies $D$ and $Q$ are used by approximately the same number of agents. From the above discussion we observe that the introduction of
strategy $\Sigma$ into the network results in two observations. Firstly, for a random network it is the dominant strategy for low Temptations. Secondly,
for a SF-like network, it stops some oscillations of the network.

\begin{figure}[!htbp]
		\centering
		\subfloat[Random 
		network]{\label{fig:rn4s_highest}\includegraphics{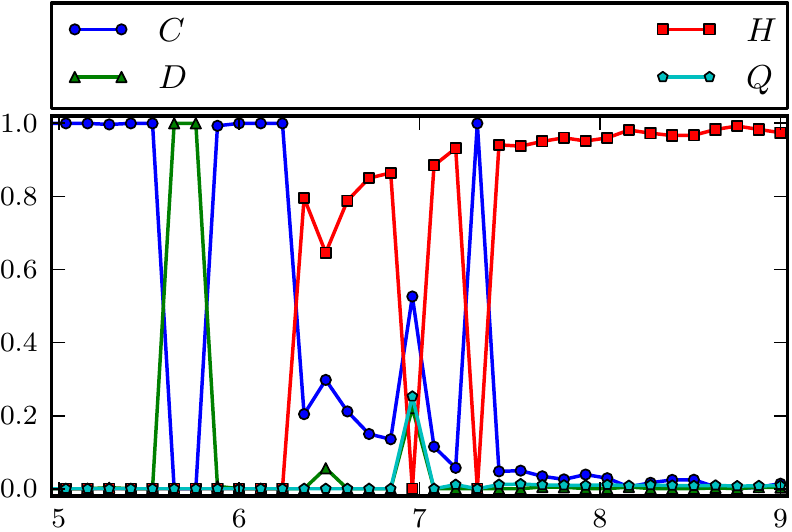}}\\
		\subfloat[SF-like 
		network]{\label{fig:sf4s_highest}\includegraphics{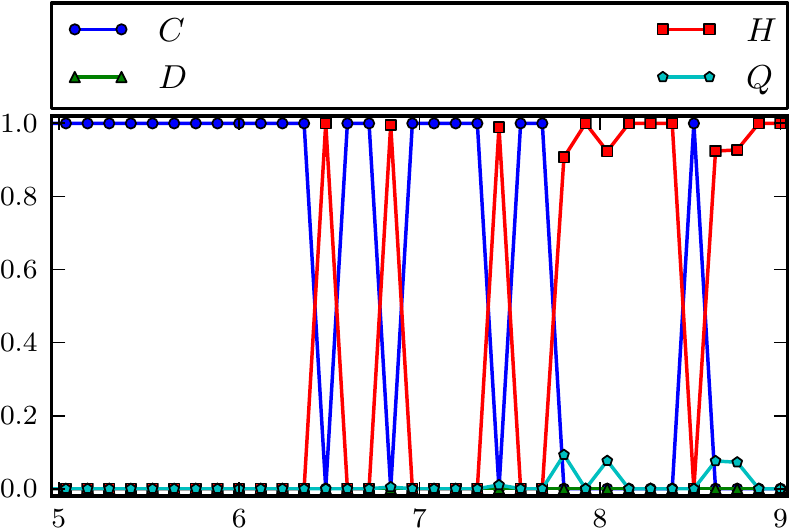}}
		\caption{Results for PD on hypergraph networks, 4 strategies, strategy $Q$ assigned to the node with highest degree.}\label{fig:4s_highest}
	\end{figure}

Next we move on to the case, where only one agent, with the highest degree was assigned a quantum strategy. For the case of four available
strategies, results are shown in Figure \ref{fig:4s_highest}. Figure \ref{fig:rn4s_highest} shows the results for a random network and
Figure \ref{fig:sf4s_highest} shows the results for a SF-like network. In this case the agent with the highest degree was assigned the $Q$
strategy, all other strategies were distributed to the agents with equal probabilities. We perceive, that for $T<6$ the strategy $C$ dominates
the network. Again, at around $T=6$ there is a shift, but this time the strategy $H$ increases its significance. The fraction of agents
using strategy $H$ slowly increases with $T$ increasing.
In the case of a SF-like network, we obtain that for $T<6.5$ the strategy $C$ dominates the network. For greater $T$ the network
starts shifting from strategy $C$ to $H$, still a small fraction of agents also use the $Q$ strategy. Summing up this case, we
can conclude that strategy $Q$ cannot infect any of the networks, but assigning the strategy $H$ to a relatively big fraction
of agents allows it to dominate the network for some values of $T$.

\begin{figure}[!htbp]
		\centering
		\subfloat[Random network]{\label{fig:rn5s_highest}\includegraphics{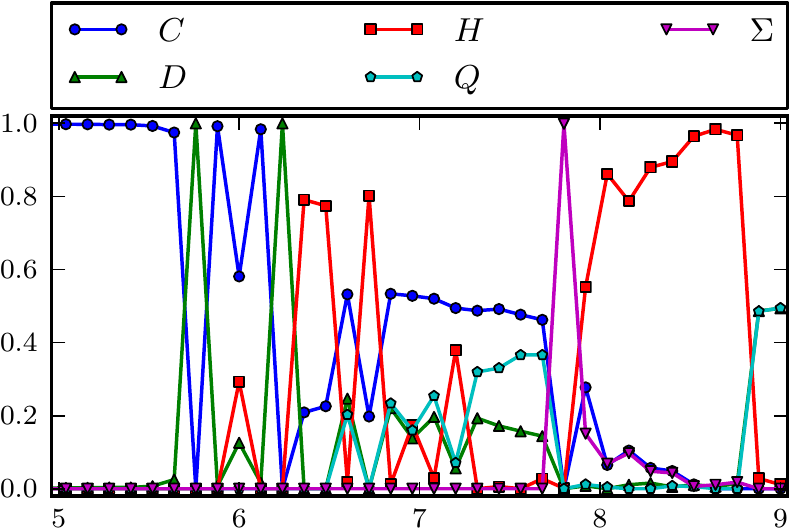}}\\
		\subfloat[SF-like network]{\label{fig:sf5s_highest}\includegraphics{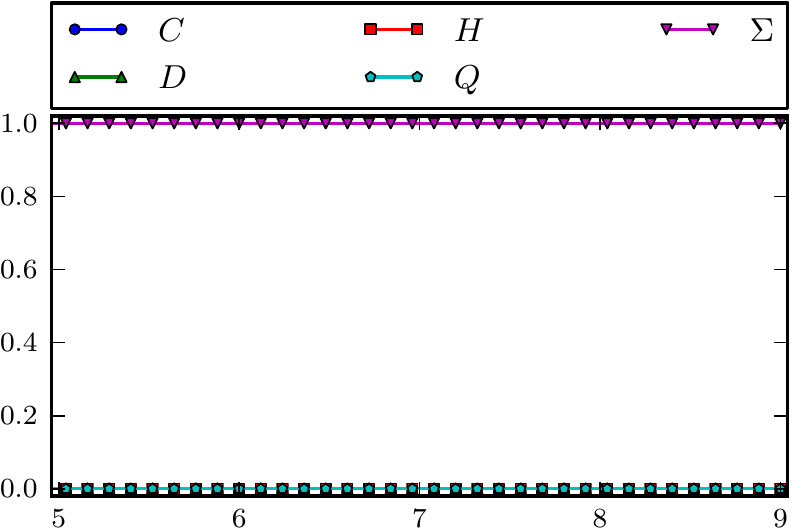}}
		\caption{Results for PD on hypergraph networks, 5 strategies, strategy $\Sigma$ assigned to the node with highest degree.}\label{fig:5s_highest}
	\end{figure}

Finally, we show the results for the case with five possible strategies. Now we assign the strategy $\Sigma$ to the agent with
the highest degree. The results obtained in this case are shown in Figure \ref{fig:5s_highest}. Figure \ref{fig:rn5s_highest} illustrates
the results for a random network and Figure \ref{fig:sf5s_highest} shows the results for a SF-like network. For a random network,
as can be seen still that for $T<6$ the $C$ strategy is employed by all of the agents. As $T$ increases from 6 to 8, the strategies
$C$, $Q$ and $D$ are used by a significant fraction of agents. At around $T=8$ the strategy $H$ dominates the network. Then,
just before $T$ reaches 9, there is another sudden shift and strategies $D$ and $Q$ are used by the same fraction of agents,
with other strategies being far less significant.
In the case of SF-like network, we observe an entirely different behaviour. The strategy $\Sigma$ always dominates the network, regardless of the value of $T$.
From this discussion it is evident that the $\Sigma$ strategy can only invade a network of a specific type. A random
network is immune to invasion.

\section{Conclusions} \label{sec:conc}

We investigate the evolution of strategies on hypergraph networks when quantum strategies $H$, $Q$ and $\Sigma$ are available
to the players. Strategies $Q$ and $\Sigma$ are considered to be invaders in our scenario. Our simulations of the evolution of strategies
on a random and SF-like hypergraph network indicate that the structure of the network is a decisive factor. In addition, we discovered that, the
strategy $\Sigma$, despite being Pareto Optimal and a Nash Equilibrium
for the three-player Prisoner's Dilemma game, does not invade the entire network in all cases. In fact, it can only invade
a SF-like network, provided that the agent with the highest degree is assigned this strategy. In other cases, depending on the value
of Temptation, the network is dominated by strategy $C$, what happens for $T<6$, or strategies $D$ and $Q$ have equal frequencies
what happens for $T>8$.
The results obtained for the case with four available strategies, are slightly different. In this case the strategy $Q$ is considered to be an
invader. The results show that a random network is invaded not by strategy $Q$, but by strategy $H$ for $T>6$. On the other hand the
SF-like network constantly shifts between $C$ and $H$ for $T>6$.

\begin{acknowledgements}
Work by J.~S{\l}adkowski was supported by the Polish National Science Center
under the project number UMO-2011/01/B/ST6/07197,
{\L}.~Pawela was supported by the Polish National Science Centre under the grant number N N519 442339.

Numerical simulations presented in this work were performed on the
``Leming'' and ``Kapibara'' computing systems of The Institute of
Theoretical and Applied Informatics, Polish Academy of Sciences.
\end{acknowledgements}

\bibliography{hypergraph}
\bibliographystyle{apsrev}

\end{document}